\documentstyle[epsfig]{aipproc}
\def\eg{{\it e.g.,\,}}
\def\ea{\ et al.\,}
\def\rel{relativistic \,}
\def\ms{\medskip}
\def\brem{bremsstrahlung}

\begin{document}                        

\title{High Energy X-Ray Emission in \\Clusters of Galaxies}

\author{Yoel Rephaeli}
\address{Tel Aviv University\\Tel Aviv, 69978, Israel}
\maketitle

\begin{abstract} Observations with the RXTE and SAX satellites have 
recently led to the measurement of a second component in the spectra of 
several clusters of galaxies which are known to have regions of extended 
radio emission. This new component is quite likely nonthermal emission 
resulting from Compton scattering of relativistic electrons by the cosmic 
microwave background. The nonthermal X-ray and radio measurements yield 
the values of the mean intracluster magnetic field, and relativistic 
electron density, for the first time in extragalactic environments. 
These results have important consequences on issues such as the origin of 
cosmic ray electrons and protons, their propagation modes in clusters, 
and the effects of these particles and magnetic fields on the intracluster 
gas. The observational results are reviewed, and some of their direct 
implications are discussed, along with near-future prospects for improved 
spectral and spatial measurements of nonthermal emission in clusters.
\end{abstract}

\section*{Introduction}

Clusters of galaxies are the largest known dynamically relaxed systems in
the universe. As such, their dynamical and hydrodynamical properties are
of basic importance not only to the study of clusters, but also to
cosmology in general. Our knowledge of clusters has been greatly
broadened by extensive X-ray measurements of thermal bremsstrahlung
emission from the hot ($10^{7}-10^{8}$ K), relatively dense intracluster 
(IC) gas which is a major component of the baryonic mass of clusters. The 
study of clusters has evolved considerably from a rudimentary stage of 
search (\eg X-ray emission) and discovery (\eg IC gas), to the more 
advanced stage of a detailed physical description of the gas, and its use 
as a probe of global cluster properties -- most importantly, the total 
mass -- and of cluster formation and evolution. X-ray measurements have
already contributed in an essential way to our understanding of the
cluster environment.

As has been the case in the study of interstellar media, an improved
knowledge of clusters necessitates the inclusion of nonthermal phenomena 
in a more precise physical description of the IC environment. The basic 
nonthermal quantities -- magnetic fields, relativistic electrons and 
protons -- may appreciably interact with the gas and thereby affect its 
thermal state. Important processes involving these fields and particles 
are radio synchrotron emission, X-and-$\gamma$-ray emission from Compton 
scattering of electrons by the cosmic microwave background (CMB) radiation, 
nonthermal \brem, decay of charged and neutral pions from proton-proton 
collisions, and gas heating by energetic protons. Detection of these 
emission yields direct information on the particles and fields, and 
establishes the basis for the study of nonthermal phenomena in clusters.

Direct evidence for \rel electrons and magnetic fields in the IC space of
clusters is provided by observations of extended radio emission in
some $\sim 30$ clusters. An appreciable IC density of \rel electrons, and 
possibly also \rel protons, implies that there could also be detectable 
X-ray and $\gamma$-ray emission in clusters (produced by some of the 
processes mentioned above). Indeed, nonthermal X-ray emission has recently 
been measured in a few clusters by the RXTE and BeppoSAX satellites. 
While we do not yet have detailed spectral and no spatial information 
on this emission, we are now able to determine the basic properties of the 
emitting electrons and magnetic fields directly from the radio and 
X-ray measurements. Spatial information will be obtained for the first 
time when the IBIS instrument aboard the INTEGRAL satellite will observe 
X-and-$\gamma$-ray emission in nearby clusters. 

In this paper I review the radio and nonthermal X-ray measurements, and 
briefly discuss some of the implications on the properties of the \rel 
electrons and magnetic fields that give rise to this emission. 

\section*{Measurements}

\subsection*{Radio Emission}

The measurement of extended low brightness IC radio emission is complicated 
by the presence of cluster radio galaxies that are typically much brighter. 
This necessitates sensitive observations over a wide range of spatial 
scales in order to carefully subtract the emission of radio sources in the 
field of view of the cluster. From a theoretical point of view, of primary 
interest is extended emission that is truly a global cluster emission; such 
emission is typically expected to be centrally positioned in the cluster. 
Extended emission associated with individual radio galaxies (\eg in lobes) 
is also of interest and can be used to probe IC magnetic fields.

In a VLA survey of 205 nearby ($z \leq 0.044$) clusters in the catalog of 
242 brightest ACO \cite{abell89} clusters detected in the ROSAT all sky 
survey \cite{ebeling96}, extended emission was detected in 29 clusters 
(see Giovannini \ea \cite{giovannini99} and references therein). Only about 
a dozen of these were previously known to have regions of extended radio 
emission. More recent VLA measurements have confirmed the presence of 
diffuse IC emission (at 1.4 GHz) in another 3 clusters \cite{giovannini00}. 
In many of the clusters the emitting region is central, with a typical size 
of $\sim 1-3$ Mpc. The emission was measured in the frequency range 
$\sim 0.04-1.4$ GHz, with spectral indices and luminosities in the range 
$\sim$1--2, and $10^{40.5}$ -- $10^{42}$ erg/s ($H_0 = 50 \;km\;s^{-1} 
\;Mpc^{-1}$). 

Magnetic field values and \rel electron energy densities can be determined 
from measurements of radio emission only if an assumption is made on 
the relative energy densities in the particles and field. Mean field values 
of a few $\mu$G were obtained under the assumption of global energy 
equipartition. The magnetic field can also be {\it estimated} from 
measurements of Faraday rotation of the plane of polarization of 
radiation from cluster or background radio galaxies (\eg \cite{kim91}). 
In the central regions of a few clusters for which rotation measures were 
determined, field values of a few $\mu$G were deduced. Note that these 
measures of the field are based on different spatial averages.

\subsection*{Nonthermal X-ray Emission}

Radio emitting IC electrons scatter off the CMB, boosting photons from this 
radiation field to the X-ray and $\gamma$-ray regions. Measurement of 
this radiation provides additional information that (when combined with 
results of radio measurements) enables the determination of the 
electron density and mean magnetic field directly, without the need to 
invoke equipartition. This was realized early on and motivated detailed 
calculations of the predicted emission \cite{r79}.

The main expected features of X-ray emission resulting from 
Compton scattering of the radio producing electrons by the CMB are: (a) 
Power-law spectrum with an index which is simply related to the radio 
index. (b) X-ray to radio luminosity ratio which is roughly equal to the 
ratio of the CMB energy density to the magnetic field energy density. 
(c) Radio spatial profile which is more pronounced than that of the 
(nonthermal) X-ray emission. Clearly, these are the predicted 
characteristics only if the X-ray and radio emissions are produced by the 
same population of electrons undergoing Compton-synchrotron energy losses. 
Note in particular that the X-ray spectral index may differ from the radio 
index if the electron spectrum is not a simple power-law. The electron 
energy spectrum may extend to energies both below and above the range 
deduced directly from the radio measurements. Low energy supra-thermal or 
trans-relativistic electrons may also give rise to power-law X-ray emission 
by nonthermal bremsstrahlung \cite{kaastra98,sarazin99}. In some of the 
proposed models for power-law X-ray emission by low-energy electrons, a 
second distinct \rel electron population is required to account for the 
radio emission. 

We have conducted the first systematic search for nonthermal X-ray 
emission in clusters by analyzing HEAO-1 measurements of six clusters 
with regions of extended radio emission \cite{r87,r88}; the search was 
continued with the CGRO \cite{r94}, and ASCA \cite{henriksen98} satellites. 
No significant nonthermal emission was detected, resulting in lower limits 
on the mean, volume-averaged magnetic fields in the observed clusters, 
$B_{rx} \sim 0.1\; \mu$G. The improved sensitivity and wide spectral range 
of the RXTE and BeppoSAX satellites have recently led to significant 
progress in the search for this emission. Clear evidence for the 
presence of a second component in the spectrum of the Coma cluster was 
seen in our analysis of RXTE ($\sim 90$ ks PCA, and $\sim 29$ ks HEXTE) 
observations \cite{r99}. While the detection of the second component 
was not significant at high energies (due to the relatively short 
HEXTE observation), we have argued that this component is more 
likely to be nonthermal, rather than a second thermal component 
from a lower temperature gas. The measurements and the two spectral 
components are shown in Figure 1. The best-fit power-law photon 
index was found to be $2.3 \pm 0.45$ (90\% confidence), in good 
agreement with the radio index. The 2-10 keV flux in the power-law 
component is appreciable, $\sim 3\times 10^{-11}$ erg cm$^{-2}$ s$^{-1}$. 
Compton origin of this component implies $B_{rx}\sim 0.2\; \mu$G, and 
an electron energy density of $\sim 8 \times 10^{-14}$ erg cm$^{-3}$ 
within a central region 1 Mpc in radius. Note, however, that we have no 
spatial information on the emission within the large field of view of the 
RXTE ($\sim 1^{o}$), so we cannot uniquely identify the source of emission. 
(Origin of the power-law emission in an active galactic nucleus does not 
seem very liekly as there is no evidence for temporal variability in the 
RXTE data.)

\begin{figure}
\centerline{\psfig{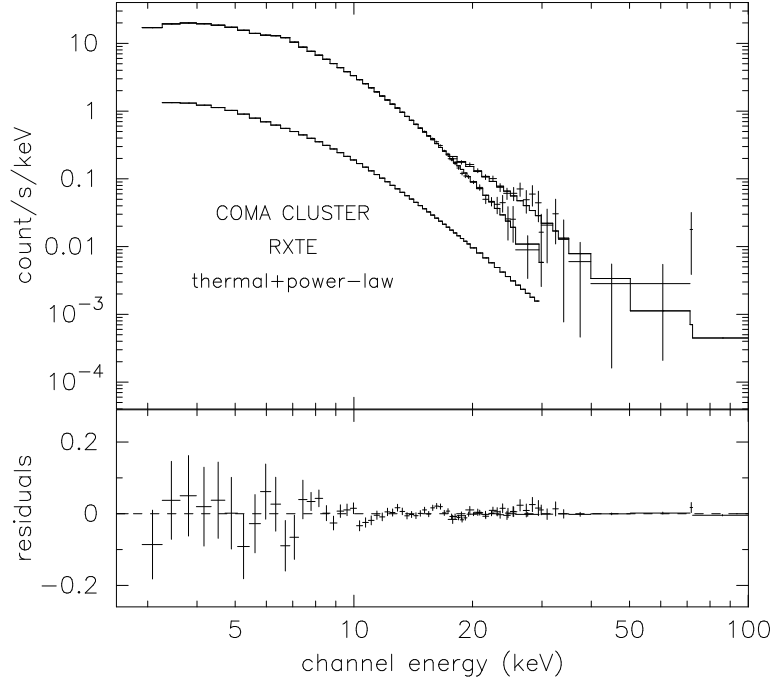}}
\caption{RXTE data and folded Raymond-Smith ($kT \simeq 7.51$ keV),
and power-law (index $=2.34$) models are shown in the upper frame; the 
latter component is also shown separately in the lower line. Residuals 
of the fit are shown in the lower frame.}
\end{figure}       

The nonthermal component in the spectrum of Coma was directly detected at 
energies $25-80$ keV when observed for $\sim 91$ ks with the PDS 
instrument aboard BeppoSAX \cite{ff99}. The PDS and lower energy 
HPGSPC data, and best-fit thermal with $kT \simeq 8.2$ keV, are shown in 
Figure 2 (Fusco-Femiano, private communication). 
A best-fit power-law photon index $2.6 \pm 0.4$ 
(90\% confidence) was deduced; the 20-80 keV flux is quite significant,  
$\sim 2\times 10^{-11}$ erg cm$^{-2}$ s$^{-1}$, and the fractional 
contribution to the 2-10 keV flux comprises $\sim 8\%$ of the total flux 
in this band, comparable to the level determined from the RXTE 
measurements \cite{r99}. A Compton origin yields $B_{rx}\sim 0.15\; 
\mu$G, close to the value we have deduced \cite{r99}.

\begin{figure}[t]
\centerline{\psfig{file=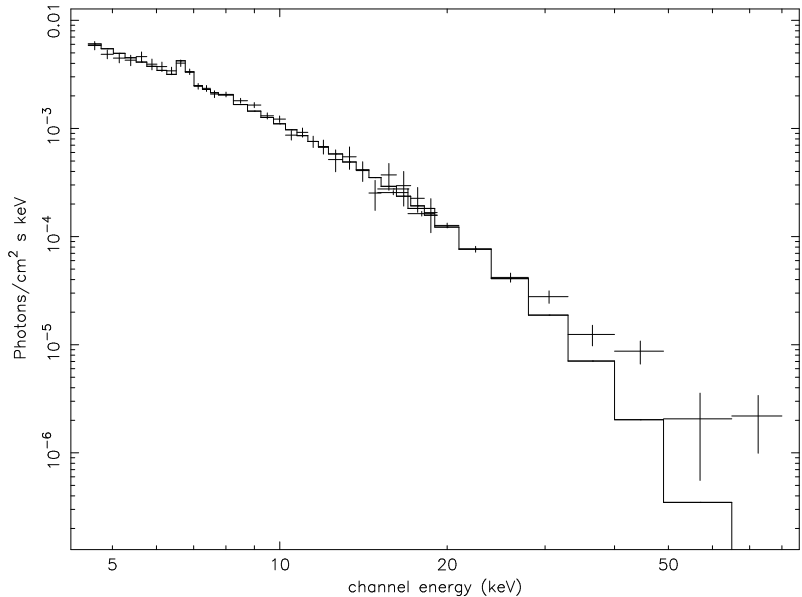, width=9cm, angle=270}}
\caption{BeppoSAX Spectrum of the Coma cluster. Shown are HPGSPC \& PDS 
data and thermal model with $kT \simeq 8.2$ keV. Significant excess 
emission is apparent in the $\sim 30-80$ keV range.}
\end{figure}    

Power-law components were detected by BeppoSAX in two more clusters, 
A2199 \cite{kaastra00}, and A2256 \cite{ff00}. Unlike Coma and A2256, 
A2199 is not known to have an extended region of IC 
radio emission, so an identification of the origin of the claimed 
detection of a power-law component in this cluster is less obvious  
than in the first two clusters. 

Emission by nonthermal electrons may possibly be detected also at lower 
energies. EUV observations of several clusters have reportedly led to the 
measurement of diffuse low-energy ($65-245$ eV) emission which is possibly 
nonthermal \cite{sarazin98,bowyer98}. However, this emission is 
said to have been unequivocally detected {\it only} in the Coma cluster 
\cite{bowyer99}.

\section*{Fields and Particles}

{\bf Fields:} Measurements of extended IC radio emission imply that the IC 
plasma is appreciably magnetized, so the first interesting issue is the 
origin of the field. While in principle the field could be of cosmological 
origin -- intergalactic seed field that had been enhanced during the 
formation and further evolution of the cluster -- it is more likely that 
the field is {\it largely} of galactic origin: IC gas is enriched in 
metals, so an appreciable fraction of it is interstellar gas from the 
member (normal and radio) galaxies that underwent ram pressure and tidal 
stripping. Since the gas is highly conductive, the field is anchored 
to the gas and ejected along with it to the IC space \cite{rcom88}. In 
this scenario it is possible to {\it qualitatively} estimate the mean 
field strength under the assumption of flux-freezing, or magnetic energy 
conservation, if re-connection is ignored. Since typical galactic fields 
are $\sim$few $\mu$G, these estimates yield a mean IC field value of a few 
$0.1 \mu$G. 

Weak IC fields could possibly be amplified by the hydrodynamic 
turbulence which is excited by galactic motions \cite{jaffe80}. 
But when a specific magneto-hydrodynamic model \cite{ruzmaikin89} for the 
growth of magnetic fields in turbulent regions was critically assessed, 
it was concluded \cite{goldman91} that the process is unlikely to 
yield mean field values which are higher than a few $0.1 \mu$G over large 
cluster regions. 

A correct measure of the mean strength of IC magnetic fields is not only 
of intrinsic importance, but is also essential in the assessment of the 
viability of \rel electron models, especially when this is based on a 
simple comparison of the Compton-synchrotron energy loss time with an 
estimated electron replenishment or acceleration time: In the analyses of 
RXTE and BeppoSAX measurements low field values were deduced (in Coma and 
A2256), as compared with the much higher values of $\sim$few $\mu$G 
that were obtained from Faraday rotation measurements. The reason for 
this -- simple, but nonetheless misunderstood by some -- becomes clear 
when one considers the difference between $B_{rx}$ and $B_{fr}$. Field 
values deduced from radio synchrotron emission depend on volume averages 
of the field over typical radial regions of $\sim 1$ Mpc, whereas rotation 
measures yield a weighted average of the field {\it and} gas density 
along the line of sight. The relation between $B_{rx}$ and $B_{fr}$ 
depends not only on the morphology of the fields -- the large-scale 
spatial variation reflecting the distribution of sources of the field, 
as well as the smaller, coherence scales ($\leq 30$) kpc over which the 
field changes direction -- but also on the gas density profile. The 
relationship between these different observational measures of the 
field were explored by Goldshmidt and Rephaeli \cite{g93}, who showed 
that $B_{rx}$ is typically expected to be smaller than $B_{fr}$. 
{\it Moreover}, the full expression for $B_{rx}$ includes a ratio of 
spatial factors (see details in \cite{r79}), which are essentially 
volume integrations of the profiles of the electrons and fields. Due 
to the lack of information on these profiles, the ratio of these factors 
has been taken to be unity in the above mentioned analyses of RXTE and 
BeppoSAX data. The effect of this simplification is {\it a systematic 
lowering} of the deduced value of $B_{rx}$.
\ms

{\bf Electrons:} The most effective energy loss mechanisms of high energy 
electrons are Compton scattering and synchrotron emission, and the 
characteristic energy loss time under these processes is a basic 
consideration in assessing models for the origin of IC \rel electrons. 
An electron emitting in the radio at frequency $\nu$ has an average Lorentz 
factor $\gamma_3 \simeq 4.9 (\nu_{8}/B_{-6})^{1/2}$, and when Compton 
scattered, it boosts a photon with the mean CMB energy ($\sim 6.3\times 
10^{-4}$ eV) to $\sim 20.2 \nu_{8}/B_{-6}$ keV ($a_p \equiv a/10^p$). 
Since Compton energy loss time (which is shorter than synchrotron 
loss time for $B < 3 \;\mu$G) is $\tau_c \simeq 2.3/\gamma_3$ Gyr, it is 
apparent that electrons which emit the observed radio emission lose their 
energy in less than 1 Gyr. If $B < 1\; \mu$G, electrons emitting at a given 
frequency have higher energies and correspondingly shorter Compton loss 
times. 

Clearly, if most of the radio-producing electrons had been injected from 
galaxies during a single, relatively short period, then the observed radio 
emission is a transient phenomenon, lasting only for a time $\sim \tau_c$ 
which is typically less than 1 Gyr. In this case the electron energy 
spectrum evolves on a relatively short timescale, and no radio emission at 
significantly higher frequencies than already observed would be expected. 
Normal cluster galaxies evolve on a longer timescale than $\sim \tau_c$, 
so that electrons from an early injection period would have lost their 
energy by now. A relatively late starburst phase \cite{atoyan00} would 
have to be invoked, perhaps resulting from galactic merger activity. 
Alternatively, electrons may have originated in a few powerful radio 
sources that had faded away.

For the observed radio and nonthermal X-ray emission to last more than 
a Compton loss time, the \rel electron population has to be sustained -- 
perhaps even in (quasi) steady state -- over timescales $>1$ Gyr. 
New electrons must be continually supplied by radio sources and other
cluster galaxies from which they diffuse (or are convected) out; 
otherwise, electrons have to be re-accelerated (in-situ) to compensate 
for their energy losses. It is not clear at all that either of these 
is a realistic possibility, and given the shorter Compton loss times 
that are naively inferred in the case of sub-$\mu$G values of $B_{rx}$, 
it has been concluded that the observed power-law X-ray emission is 
unlikely to be due to Compton scattering of the same electrons that 
produce the radio emission. At electron energies much lower than $\sim 1$ 
GeV, the main energy loss process is electronic (or Coulomb) excitations 
\cite{r79}; as a result, the total loss time is maximal for $\gamma \sim 
300$ \cite{sarazin99}. For electrons with energies near this 
value to produce the observed IC radio emission, the mean magnetic field 
has to be at least $\sim$few $\mu$G in strength. And while such electrons 
could possibly produce EUV emission by Compton scattering off the CMB, 
their energies are too low to boost CMB photons to energies $> 1$ keV.

A quasi steady state may also be reached if the electrons are continually 
accelerated by a first order Fermi acceleration mechanism in shocks 
(generated, perhaps, during galactic or subcluster mergers), if this 
process is effective also in the typically trans-sonic IC shocks.
Models of accelerating electrons have been proposed (\eg 
\cite{kaastra98,sarazin00}) in which the electrons produce (nearly) 
power-law X-ray emission by nonthermal bremsstrahlung. 
The range of electron energies and values of the power-law indices 
that are consistent with the radio, EUV, and X-ray measurements in the 
Coma cluster and A2199 have been determined by Sarazin and Kempner 
\cite{sarazin00}. It is claimed that parameters can be selected such 
that accelerating electron populations can marginally produce the 
observed radio and power-law X-ray emission, but not the EUV emission. 
However, for the range of parameters for which consistency is obtained, 
the electron energy density is very high, much higher than that of the 
\rel electron population that produces power-law X-ray emission by 
Compton scattering. These specific accelerating models seem therefore 
to be less likely on energetic grounds.
\ms

{\bf Protons:} In galaxies, protons are the main cosmic ray component, and 
it is expected that this is so also in clusters. The low energy proton 
density can be determined if $\gamma$-ray emission from neutral pion 
decays is observed, or if the radio and X-ray producing electrons are 
secondaries resulting from charged pion decays (as suggested in 
\cite{blasi99}). These processes were first considered in a cluster 
context by Dermer and Rephaeli \cite{d88} who estimated the high energy 
spectrum of M87 and the Virgo cluster taking into account all relevant 
processes. If the proton-to-electron ratio in the IC space is not much 
smaller than its Galactic value, then energy deposition into the gas by 
low energy protons (through Coulomb interactions) may also be important 
and could affect the thermal state of the gas, particularly in a central 
cooling flow region \cite{r95}.

\section*{Future Prospects}

Measurements of diffuse IC radio emission, and the initial detection of 
nonthermal X-ray emission in three clusters, provide strong 
motivation for further radio and X-ray observations. In addition to 
the need to extend the search for diffuse radio emission in many other 
clusters, it is very important to measure the spectrum over a wider 
frequency range in order to determine spectral indices and the 
implied range of \rel electron energies.

We are continuing the search for nonthermal X-ray emission with the 
RXTE; the cluster A2319 was recently observed for $\sim 160$ ks, and a 
second, longer ($\sim 200$ ks) observation of the Coma cluster is 
scheduled for late 2000. With the recent observation of A1367 by 
BeppoSAX, we will soon have the results from (at least) 5 clusters 
which have been observed by these two satellites with the main purpose 
of searching for nonthermal emission. 

The next important step in the exploration of cluster nonthermal 
X-ray emission will be taken by ESA's INTEGRAL satellite, scheduled to 
be launched in 2002. Of particular interest are the capabilities of the 
IBIS imager, which will cover the spectral range of 15 keV -- 10 MeV, and 
it will -- for the first time -- provide moderate spatial resolution down 
to $\sim 12$ arcminute (FWHM). A very deep exposure ($\sim 1000$ ks) 
of a {\it nearby} cluster will likely enable spatial mapping of nonthermal 
emission in the energy range $\sim 30 - 100$ keV, if the cluster emission 
is at the level measured by RXTE and BeppoSAX. The spatial information will 
be crucial for identifying the origin of the nonthermal emission, and 
a comparison between the X-ray and radio morphologies could possibly 
provide detailed information on the \rel electrons and magnetic field 
spatial distributions. Moreover, the improved spectral information that 
is expected from observations with IBIS will likely enable more definite 
deductions on the electron energy spectrum, and the main nonthermal 
X-ray emission mechanism.

\section*{Acknowledgments}

I am grateful to Dr. Duane Gruber for our long collaboration on the 
analysis of HEAO-1, CGRO, and RXTE measurements of clusters, and to 
Dr. Roberto Fusco-Femiano who kindly provided the BeppoSAX spectrum 
of the Coma cluster (Figure 2).

\end{document}